\documentclass[twocolumn,showpacs,preprintnumbers,amsmath,amssymb]{revtex4}

\usepackage{graphicx}
\usepackage{dcolumn}
\usepackage{bm}

\begin{document}

\title{Reactive glass  and  vegetation patterns}

\date{\today}

\author{ N.M. Shnerb$^1$ , P. Sara$^2$ ,H. Lavee$^2$ and S. Solomon$^3$}
\affiliation{$^1$Department of Physics, Judea and Samaria College, Ariel, Israel 44837}
\affiliation{$^2$ Department of Geography, Bar-Ilan University, Ramat-Gan, Israel 52900}
\affiliation{$3$ Racah institute of physics, The Hebrew University, Jerusalem Israel 91904}

\begin{abstract}
The formation of vegetation patterns in the arid and the semi-arid climatic zones is studied. Threshold for the
biomass of the perennial flora is shown to be a relevant  factor, leading to a frozen disordered patterns in the
arid zone. In this ``glassy'' state,  vegetation appears as a singular plant spots separated by irregular
distances, and an indirect repulsive interaction among shrubs is induced by the competition for water. At higher
precipitation rates, the diminish of hydrological losses in the presence of flora becomes important and yields
spatial attraction and clustering of biomass. Turing-like patterns with characteristic length scale may emerge
from the disordered structure due to this positive feedback instability.
\end{abstract}

\maketitle

Vegetation patterns in the arid and the semi-arid climatic zones
\cite{tbs1,tbs2} are an interesting example of spontaneous
symmetry breaking in complex
systems. Competition of shrubs for a limited supply of water
is the relevant process that dictates the spatial organization.
The struggle for water induces an indirect
interaction among shrubs, as the flora extinct
if its water supply is insufficient.

Competition for common resource has been considered for many years as one of the basic processes in population
dynamics \cite{gause,tilman}. It may be shown that, if two species compete for a common resource, the one that
is able to survive at a lower resource level prevails and displaces the other species population. Stable
coexistence of N species is possible if there exist N different resources and each of the species is a superior
competitor for one of the supplies, that is, it has one \it biological niche. \rm

The situation becomes more complicated if the resource admits spatial dynamics. Recent theoretical and
experimental work reveals the dynamics of competing populations in water, where light, the limiting resource, is
consumed gradually by the upper layers of aquatic phytoplankton \cite{huisman}. This model may be extended to
include spatial dynamics of the fauna,  but it does not support
 time independent  patterning .

Vegetation patterns are an example of one species (shrubs)
and one resource (water) system, where field studies revealed wide
variety of stable, or almost stable,
spontaneous segregation modes. Understanding the underlying
mechanism for generation of such patterns and their observed resilience
is considered as an important step toward a comprehension of the
desertification
process, where environmental effects like climate changing and grazing
destroys the natural balance toward stable aridity.

Technically, the water-biomass system has been considered as a spatially extended nonlinear system, that, in
some parameter range, may yield stripes, spots, labyrinth and other ordered arrangements attributed to a
positive feedback mechanism, i.e., to the inhibition of water runoff and evaporation by the flora \cite{wilson}.
However, the typical perennial vegetation patterns in the semi-arid zone are disordered, as one can easily see
in Fig. (1). The generic spatial organization of perennial flora varied along the precipitation gradient: from
scattered "green spots" in the arid zone through clusters of shrubs in the semi-arid zone, to an almost full
coverage of the soil by biomass in the humid/subhumid climate. Analysis of the transverse correlations  in the
three panels of Fig. (1)  shows that the correlation length in the semi-arid zone is larger (by factor of 2-3)
than in the other regions, and seems to indicate weak long-range oscillations in the Mediterranean site, perhaps
a precursor of Turing instability.

\begin{figure*}
\includegraphics{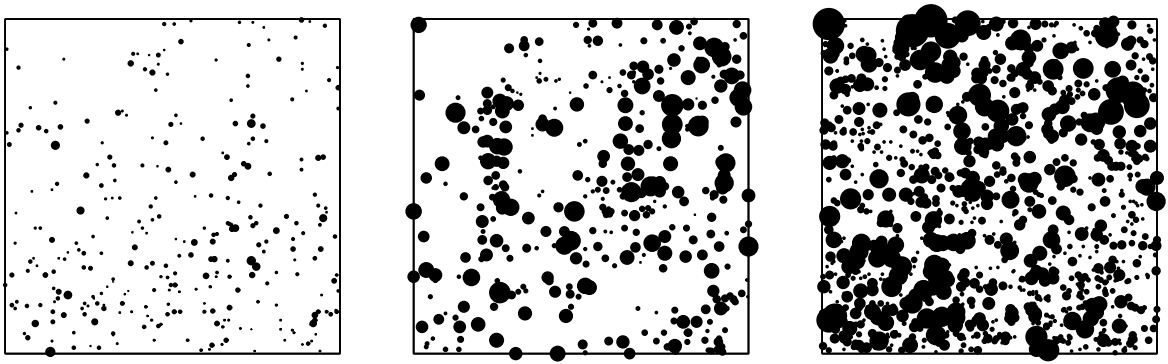}
\caption{\label{fig1} Results of direct  field measurements
at three different locations
along the precipitation gradient.
The distribution of perennial
shrubs (annual flora not included)  is presented
for an  area of 100 square meters at each site.
Each  black spot presents a
shrub and the size of a
spot is proportional to the area size
of the canopy. Shrubs
distribution on hillslopes has been taken  at three sites representing
mildly-arid, semi-arid and
subhumid  climate
conditions in Israel.
The mildly-arid site (left
panel, with mean annual rainfall 260mm, hillslope
gradient $11^o$) is located at   Mishor
Adumim, 10km east of Jerusalem.
The semi-arid
site
(middle, with  mean annual rainfall 330mm and gradient
of $16^o$), is located at
Ma'ale Adumim, 8km east of
Jerusalem.
The Mediterranean site (right, mean annual rainfall  620mm,
hillslope gradient $13^o$), is located at
Giv'at Ye'arim, 11km west of Jerusalem.
All the three sites  have hard
calcareous bedrock and southeast exposure (azimuth  $140^o-150^o$).}
\end{figure*}

In this letter we present a general and simple model of the water-shrubs
reaction that is able to yield all these features. Our model takes
into account the intrinsic ``noise'', i.e., the amplification
of initial fluctuations due to the minimal size needed for the survival
of perennial flora. The resulting pattern is disordered but robust,
thus it may be considered as the reactive equivalent of a glass. Although
the concept of free energy, deep local minima and thermal equilibrium
is absent in reactive systems, it still presents an example of a spontaneous
breakdown of symmetry toward a disordered, long-living meta-stable
state.

To present the model, let us begin from its zero
dimensional  ("flower pot") deterministic
and continuous dynamics. With water supplied to the system at some
rate \( R \) and continuum vegetation growth, the time evolution
of the water-shrub system is described by the following,
nondimensionalized rate equations:

\begin{eqnarray}
\label{1}
\frac{\partial B}{\partial t}&=&wB-\mu B\\   \nonumber
\frac{\partial w}{\partial t}&=&R-w-\lambda wB
\end{eqnarray}

Where \( w \) stands for the available water density, B is the density
of shrubs biomass, the term \( wB \) represent shrubs growth as they
consume water while \( -\lambda wB \) is the corresponding consumption
of water by shrubs. \( \mu  \) is the \char`\"{}death rate\char`\"{}
of the vegetation in the absence of water and the term \( -w \) represents
water losses by percolation  and evaporation.

The set of differential equations (\ref{1}) admits two non-negative
fixed points. The trivial one at \( B_{0}=0,\: w_{0}=R \), becomes
unstable to small perturbations as \( R=\mu , \) while above this
value a ``coexistence'' fixed point at
\begin{equation}
\label{2}
w_{1}=\mu \quad B_{1}=\frac{R-\mu }{\lambda \mu },
\end{equation}
becomes a stable node (if \( R^{2}<4(R-\mu )\mu ^{2} \)) or a stable
focus (if \( R^{2}>4(R-\mu )\mu ^{2} \)).

Adding lateral water flow to the above model leads to a reaction-diffusion
equations of the form,

\begin{eqnarray}
\label{3}
\frac{\partial B}{\partial t}&=&wB-\mu B\\ \nonumber
\frac{\partial w}{\partial t}&=& D \nabla ^{2}w+ {\bf v}
\cdot \nabla w+R-w-\lambda wB,
\end{eqnarray}
where \( \bf{v} \) pointed down the hillslope. Simple linear analysis
implies that in the absence of cross-diffusion effects (like those
considered recently  by \cite{meron}), no Turing-like instability exists in
that system; the steady state is a uniform covering of all the plane
by the same amount of flora which corresponds to the stable fixed
point and  fluctuations of wavenumber \( k \) decay like \( e^{-k^{2}t} \).

In the desert area considered here there are two seasons, dry
summer and humid winter. Eqs. (\ref{2},\ref{3}) presents the
winter, with ``smeared'' rain events. While annual flora wilt in
the summer, perennial shrubs have to  survive, so they must  reach
some \textit{threshold size} before the dry season. If the winter
is not long enough to allow for a full development of the plant to
its stable fixed point, the survival of a shrub depends on its
size at the end of the rainy season, which, in turn, depends on
small fluctuations in its initial size and the consumption of
water by its neighbors.

In the next winter, the existence of a shrub causes a depletion of
the available soil moisture in its immediate neighborhood (roughly
speaking, in an area of typical linear size \( \sqrt{1/\mu } \)),
and the chance for another shrub that pops out in the depletion
region to reach the threshold is lowered. The whole area is then
segregated into a mosaic of water accepting and water contributing
patches. This is a non-Turing mechanism  that has nothing with the
effect of shrubs on the overland flow. While Turing instability is
characterized by some typical wavelength that sets the linear size
of vegetation and bare soil patches to be equal, our model  allows
for clusters of arbitrary size, as indicated in Fig. (1).

The optimal segregation of the hillslope,
that gives maximal biomass per unit area, is an ordered
array of shrubs, each located in the lower end (or, on a
flat plane, in the middle) of its contributing area. A
regular or distorted  "lattice" of flora
is  then formed, similar to the structure of atoms in a two dimensional
crystal. However, this optimal scenario is rarely accomplished in nature,
due to the stochastic character of the growth process itself. For
simplicity, let us assume that the seed bank in the soil ensures the
development of a perennial shrub if some water exist at the site.
As the first shrub pops out in an empty region, the soil moisture in
its surroundings (primarily downhill) is depleted, and the next shrub
will not grow in this "shadowed" area. Nevertheless,
as the next shrub also occurs at random, its position will be uncorrelated
with the first, except that it can not pop out at the shadowed region
of the first one, and so on. The process continues until all the
slope is shaded (``jammed''). This stochastic growth process yields a random
covering of the slope by shrubs, with typical distance between nearest
neighbors, but no long-range structure. The random arrangement is,
however, extremely robust; although the slope is covered inefficiently
by the shrubs, there is not enough source area for the next shrub.
Essential changes, such as the death of a plant
and a formation of another one, are discontinuous, and though
are very rare, unless some intervene come from the outside, perhaps
in the form of grazing or climate changes.

\begin{figure}
\includegraphics[width=6cm]{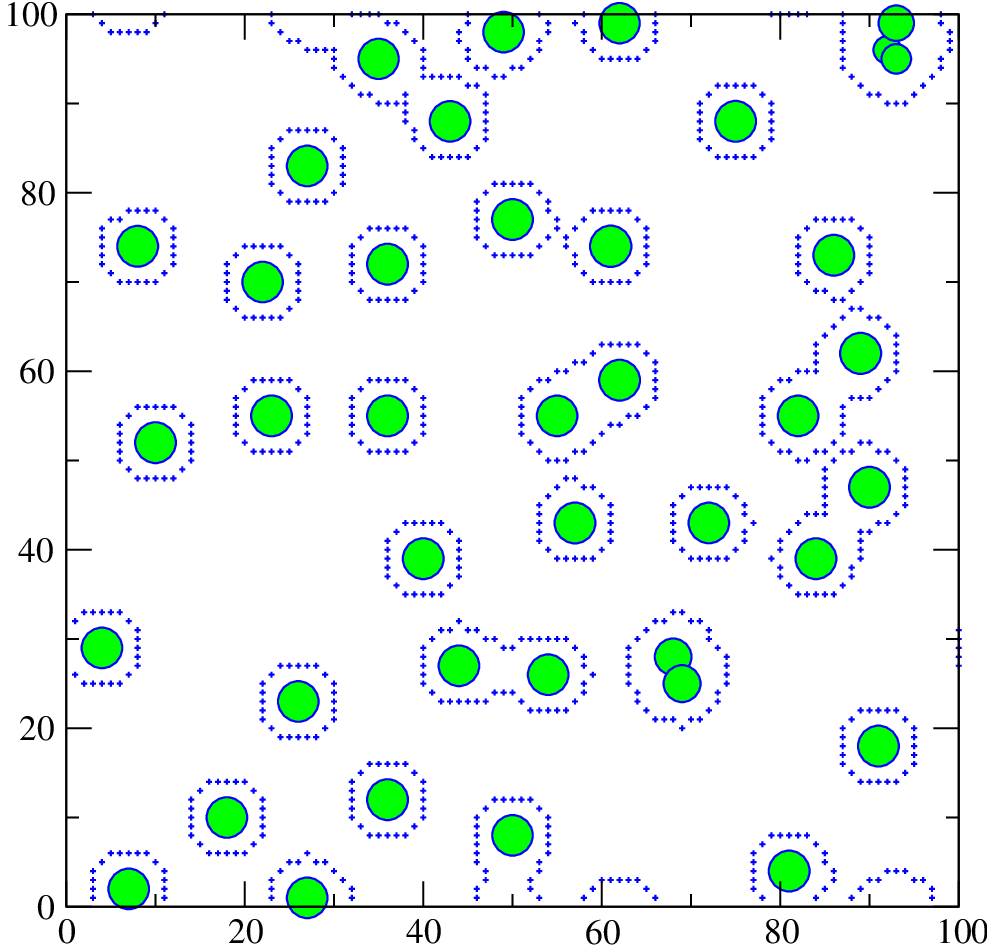}
\caption{\label{fig2}Numerical results of forward  Euler
integration of the reaction-diffusion equations (\ref{3}) on
100X100 sites grid with periodic boundary conditions. A grey spot
is plotted around the location of each shrub, with  the size of
the spot proportional to its biomass. The dotted lines present
soil moisture contours around the shrubs. The simulation
parameters are \( v=0 \) (no slope) and $ D=10$, \( \mu =0.2,\:
R=0.5,\: \lambda =1.2 \) (this implies \( B_{1}=1.25 \) and \(
w_{1}=0.2 \)). Initial conditions are no water and a seed of
biomass taken from a square distribution between {[}0,0.01{]}.
Effect of the summer is modelled at the end of each ``winter'' (21
time cycles) by setting all water to zero while the flora at a
site is dropped to the seed level if the biomass is smaller than a
threshold, with \( B_{th}=1.2 \). The average (per site) values of
water (about 0.32)  and flora (0.57) reflect the inefficient use
of water attributed to the glass-like structure.}
\end{figure}

\begin{figure}
\includegraphics[width=6cm]{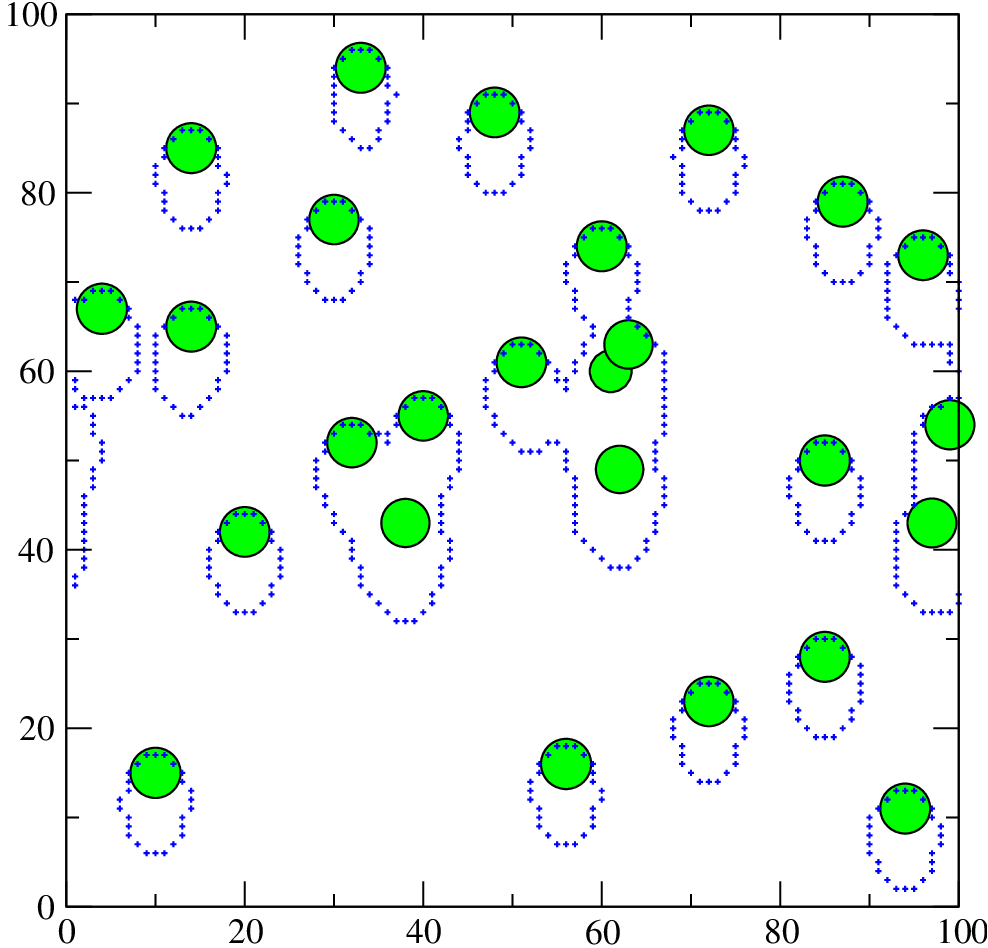}
\caption{\label{fig3}
Same as Fig. (\ref{fig3}) but with a downhill drift of water
where the asymmetry parameter is $v=0.5$. Note the change in the shape of
the water contours. Average biomass and water levels are almost the same as
in Figure (\ref{fig2}).}
\end{figure}

In Figure (\ref{fig2}), typical results of a numerical simulation
of a threshold-noise model on a ``flat'' (no slope) land  are
presented.  In the simulation, the system freezes rapidly (in 1-2
rainy seasons) to a robust  configuration (slower dynamics will
increase the efficiency of biomass growth), and then   persists,
with negligible  fluctuations, up to 500 winters. The average
amount of flora is much smaller than \( B_{1}, \) and the average
amount of water is larger than \( w_{1} \), i.e., there is an
inefficient use of the water due to the stochastic arrangement of
the shrubs. To guide the eye, soil moisture contours  are also
plotted (dotted lines, with the water level inside  smaller than
0.46) and reveal the depletion zone around each shrub \cite{note}.
No empty site in the region maintains enough water to allow for
new shrub to develop, and the whole region is ``shaded'' by the
existing flora. Figure (\ref{fig3}) present the results of a
simulation with the same growth and diffusion  parameters, but
with   nonzero  downhill slope, and  the  effect on the moisture
depletion zones is evident.

\begin{figure*}
\includegraphics{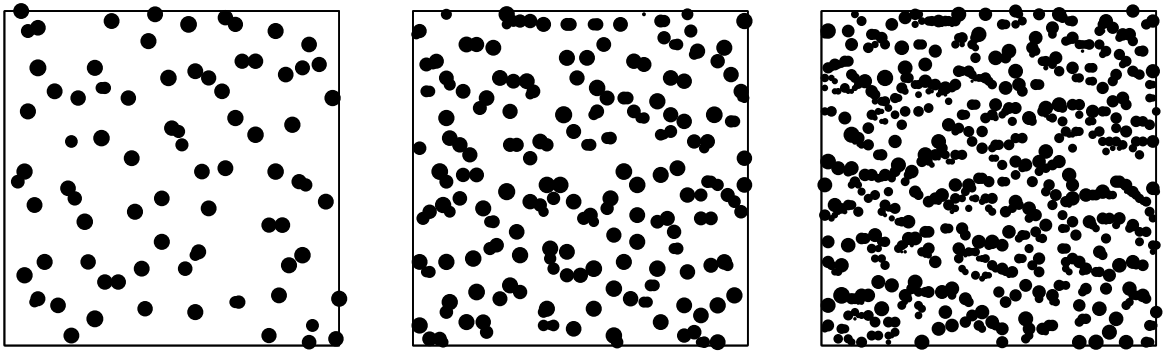}
\caption{\label{fig4}
Effect of the  positive feedback mechanism.  While
at low precipitation ($R=0.45$, left panel) the repulsive interaction wins
and no clustering occurs, at higher humidity (R=$0.47$, middle panel)
clustering is more
pronounced, and at $R=0.5$ (right)
vegetation stripes may be recognized. All other
parameters are the same as in Figure (\ref{fig3}).}
\end{figure*}

Various aspects of this competition scenario are similar to the
adsorption  of large particles at an interface \cite{tbs4}. In the
model of random sequential adsorption, "hard" particles are added
sequentially to a D dimensional volume at random positions with
the condition that no trial particle can overlap previously
inserted ones. The addition process is than repeated until the
system reaches its "jamming limit", at which the density
saturates. The adsorbed particles density at the jamming limit is
lower than the close-packed form, and the configuration of
adsorbed particles  is "freezed" at some disordered pattern. A
shrub above threshold, with its excluded volume of depleted
moisture is similar to an adsorbed ``disc''. The shrubs-water
model,  although  non-local (excess water transferred downhill)
and reversible (new shrub may  remove an existing plant by
depleting its water resources), yields  a similar jamming
disordered and inefficient covering of the slope.

The above considerations about glassy structure elucidates the
existence and the robustness of the vegetation patterns in the
arid zone, but fails to explain the aggregation of shrubs (larger
correlation length) in the semi-arid zone, or the Turing patterns
\cite{wilson,meron}. To explain patchiness one should consider the
positive-feedback mechanism, i.e., the inhibition of water
dynamics induced by the shrubs themselves. In the absence of
shrubs (and other meso topography factors), water flow downhill,
with some typical lateral displacement per unit length. In the
presence of shrubs the flow in their vicinity is slower than that
in bare soil, as a result of higher infiltration rates. In
addition the microclimate under shrub is characterized by less
direct radiation and  smaller evaporation rate \cite{wilson}. This
means that close to the shrub there are favorable soil water
conditions and more flora may grow. The ``repulsive'' interaction
among shrubs due to the struggle for water is then balanced by an
``attraction''. Accordingly, the size of a typical cluster changes
along the precipitation gradient, from a single shrub at the arid
limit to large clusters in the semi-arid regime. The response of
the system to external parameters (climate change) and grazing
seems to depend on its phase. In particular, it seems that
hysteresis loops (desertification transition) like those described
in \cite{meron} are a characteristic of the clustered phase. A
detailed discussion of these issues will be presented elsewhere.

In Fig. (\ref{fig4}), the results of the same simulation program with positive feedback are presented. The only
new ingredient added to the simulation is a suppression of the asymmetry in the downhill water flow, with the
asymmetry term $\bf{v}$
 multiplied by $exp[-5B({\bf r})/B_{th}]$ (this is a cross-convection effect). Diffusion
and evaporation remain the same as in (\ref{3}). As the downhill flow becomes smaller, water tend to accumulate
at shrub's neighborhood.  The transition from the arid (left) zone, with no clustering and glassy structure, to
more clustered patterns and even some linear order in the semi-arid (right) is evident.

In conclusion, the generic  spatial patterns due to the struggle for water are disordered  frozen patterns, and
the threshold-noise process dictates the vegetation spatial organization in the arid zone. The instability that
yields this glassy structure is not Turing-like, thus the inhomogeneity is not characterized by typical length
scale. Shrub clusters and  Turing patterns  emerge as further instability of  this glassy structure if the
precipitation is large enough, where the positive feedback dominates. It  then leads first to clustering of
shrubs and then to a global order in the form of "tiger bush" patterns.

\begin{acknowledgments}
We wish to acknowledge  E. Meron, F.J. Weissing
and J. Huisman for most helpful comments and discussions.
\end{acknowledgments}


\begin{thebibliography}{}
\bibitem{note} Actual field experiments usually finds the soil moisture
under a shrub larger than the moisture few meters away from it. The reason
for that has to do with the positive feedback mechanisms that reduce water
losses in this region. However, up to some limit, this water resource serves
the shrub during the summer, and is not available for the use of another
planet. One may regard this fraction  of soil moisture as part of the biomass,
while the contours in  Figs. (\ref{fig2},\ref{fig3},\ref{fig4})
represent available water density.
\end{thebibliography}
\end{document}